\documentclass[english]{article}
\usepackage[T1]{fontenc}
\usepackage[latin9]{inputenc}
\usepackage{amsbsy}

\makeatletter
\date{}

\makeatother

\usepackage{babel}

\begin{document}

\title{A note on \\
multi-time equations in quantum mechanics}

\author{{\normalsize Ali Sanayei}\\
\emph{\small Institute for Theoretical Physics, University of T\"ubingen}\emph{
}\\
\emph{\small Auf der Morgenstelle 14, 72076 T\"ubingen, Germany}\emph{}\\
{\footnotesize Email: ali.sanayei@uni-tuebingen.de}}
\maketitle
\begin{abstract}
The main aim of this note is to show that the formalism of multi-time
equations in quantum mechanics meant to represent a manifestly Lorentz-invariant
theory suffers from at least three imperfections: (1) It does not
cover those physical systems for which the Schr\"odinger picture of
quantum mechanics does not exist. (2) The integrabiliy condition which
has been proposed in the formalism does not imply as necessary that
for every partial Hamiltonian there exists a Schr\"odinger equation.
(3) The formalism does not generally allow for interactions even in
a non-relativistic multi-particle quantum system.
\end{abstract}
Recently, a formalism of multi-time equations involving several time
variables, one for each particle, has been discussed {[}11{]} so that
quantum mechanics can manifestly be Lorentz invariant. It was claimed
that the relativistic analogue to a non-relativistic wave function
$\psi\left(t;x_{1},\ldots,x_{n}\right)$ is a multi-time wave function
$\phi\left(t_{1},x_{1};\ldots;t_{n},x_{n}\right)$, on a set of space-like
configurations, for which $\phi=\psi$ when $t_{1}=\cdots=t_{n}=t$
{[}11{]}. Furthermore, it was claimed that the time evolution of $\phi$
is governed by the Schr\"odinger equation

\begin{equation}
i\partial_{t_{j}}\phi=\mathrm{\hat{H}}_{j}\phi\end{equation}
for all $j=1,\ldots,n,$ with partial Hamiltonians $\mathrm{\hat{H}}_{j}$
which (at least when all $t_{j}$ are equal) add up to the full Hamiltonian
$\mathrm{\hat{H}}$ fulfilling the Schr\"odinger equation

\begin{equation}
i\partial_{t}\psi=\mathrm{\hat{H}}\psi.\end{equation}
Moreover, the offered formalism refers to an integrability condition

\begin{equation}
\frac{\partial\mathrm{\hat{H}}_{k}}{\partial t_{j}}-\frac{\partial\mathrm{\hat{H}}_{j}}{\partial t_{k}}+i\left[\mathrm{\hat{H}}_{j},\mathrm{\hat{H}}_{k}\right]_{-}=\hat{0}\end{equation}
and to a Stone-based theorem, whereby the partial Hamiltonians were
assumed to be self-adjoint (see {[}11{]} and references therein).
It was claimed that the system of equations (1) will then be consistent. 

The aim of the present note is to show three points: (i) The formalism
does not cover those physical systems for which Schr\"odinger picture
of quantum mechanics does not exist. (ii) It is feasible to indicate
a kind of decomposition of the full Hamiltonian for which (3) holds,
however, (1) does not hold. (iii) Hence, the formalism only admits
particular decompositions of a full Hamiltonian and even in this case
does it not take into account interactions in the case of a non-relativistic
multi-particle quantum system.

First, since the whole problem has been formalized in the Schr\"odinger
picture, it is perhaps worth mentioning that Schr\"odinger picture and
hence a Schr\"odinger equation governing a wave function do exist when
the corresponding Hamiltonian is Hermitian. But it is possible to
indicate physical systems for which the Schr\"odinger picture does not
exist although the Heisenberg picture exists (see {[}7,8,9{]}). Such
physical systems usually refer to those containing an infinite number
of degrees of freedom {[}7,8{]}. More precisely, there are physical
systems whose Hamiltonians together with boundary conditions are not
Hermitian and therefore there is no Schr\"odinger equation to govern
them. For instance, Gamow states which are wave functions of resonances
are eigenvectors of a Hamiltonian possessing complex eigenvalues (e.g.,
see {[}10{]}). The dispersion phenomenon for nuclear reactions (Siegert
states) {[}4{]} where the Hamiltonian together with the boundary conditions
is not self-adjoint would be another example. A state vector of those
systems moves about in a some more general space called the rigged
Hilbert space. Thus, due to the fact that the basis of the formalism
is the Schr\"odinger picture, it will automalically miss several real
physical systems for which the Schr\"odinger picture does not exist.

Second, let us consider decomposition of a Hilbert space by direct
sums. Thus, assume a Hermitian Hamiltonian $\mathrm{\hat{H}}$ (which
is not an explicit function of time) in a full Hilbert space $\mathcal{H}$
with two subspaces $\mathcal{A}$ and $\mathcal{B}$, where $\mathcal{B}=\bar{\mathcal{A}}=\mathcal{H}-\mathcal{A}$.
Since $\mathrm{\hat{H}}$ is Hermitian, there exists a Schr\"odinger
equation governing it. On the other hand, since $\mathcal{H}$ consists
of two subspaces, the Schr\"odinger equation for $\mathrm{\hat{H}}$
can be represented as a matrix equation

\begin{equation}
\left[\begin{array}{cc}
\hat{\mathrm{H}}_{\mathcal{AA}} & \mathrm{\hat{H}}_{\mathcal{AB}}\\
\mathrm{\hat{H}}_{\mathcal{BA}} & \mathrm{\hat{H}}_{\mathcal{BB}}\end{array}\right]\left[\begin{array}{c}
\left|\psi_{\mathcal{A}}\right\rangle \\
\left|\psi_{\mathcal{B}}\right\rangle \end{array}\right]=i\partial_{t}\left[\begin{array}{c}
\left|\psi_{\mathcal{A}}\right\rangle \\
\left|\psi_{\mathcal{B}}\right\rangle \end{array}\right],\end{equation}
where $\mathrm{\hat{H}}_{\mathcal{AA}}=\mathrm{\hat{P}}_{\mathcal{A}}\mathrm{\hat{H}}\mathrm{\hat{P}}_{\mathcal{A}}$,
$\mathrm{\hat{H}}_{\mathcal{BB}}=\mathrm{\hat{P}}_{\mathcal{B}}\mathrm{\hat{H}}\mathrm{\hat{P}}_{\mathcal{B}}$,
$\mathrm{\hat{H}}_{\mathcal{AB}}=\mathrm{\hat{P}}_{\mathcal{A}}\mathrm{\hat{H}}\mathrm{\hat{P}}_{\mathcal{B}}$,
$\mathrm{\hat{H}}_{\mathcal{BA}}=\mathrm{\hat{P}}_{\mathcal{B}}\mathrm{\hat{H}}\mathrm{\hat{P}}_{\mathcal{A}}$,
$\mathrm{\hat{P}}_{\mathcal{A}}+\mathrm{\hat{P}}_{\mathcal{B}}=\hat{1},$
and $\mathrm{\hat{P}}$ denotes the orthogonal projection operator.
It is straightforward then by (4) to show that:

\begin{equation}
\left(\mathrm{\hat{H}}_{\mathcal{AA}}+\frac{\mathrm{\hat{H}}_{\mathcal{AB}}\mathrm{\hat{H}}_{\mathcal{BA}}}{i\partial_{t}-\mathrm{\hat{H}}_{\mathcal{BB}}}\right)\left|\psi_{\mathcal{A}}\right\rangle =i\partial_{t}\left|\psi_{\mathcal{A}}\right\rangle \end{equation}
Although the full Hamiltonian $\mathrm{\hat{H}}$ in the Hilbert space
$\mathcal{H}$ fulfills the Schr\"odinger equation (2), (5) does not
itself represent the Schr\"odinger equation for the partial Hamiltonian
$\mathrm{\hat{H}}_{\mathcal{AA}}$ because there exists a further
perturbative term. Notwithstanding the fact that the integrability
condition (3) holds, that is,

\begin{equation}
\left[\mathrm{\hat{H}}_{\mathcal{AA}},\mathrm{\hat{H}}_{\mathcal{BB}}\right]_{-}=\left[\mathrm{\hat{P}}_{\mathcal{A}}\mathrm{\hat{H}}\mathrm{\hat{P}}_{\mathcal{A}},\mathrm{\hat{P}}_{\mathcal{B}}\mathrm{\hat{H}}\mathrm{\hat{P}}_{\mathcal{B}}\right]_{-}=\hat{0},\end{equation}
(1) does not necessarily hold for every partial Hamiltonian. If one
takes Dirac-Frenkel variational principle {[}1,2{]} as a numerical
approximation method whereby one assumes that the same full Hamiltonian
$\mathrm{\hat{H}}$ holds in a subspace $\mathcal{V}$ of a full Hilbert
space $\mathcal{H}$, even in this case there exists an error operator
between the time derivative obtained in $\mathcal{H}$ and the time
derivative obtained in $\mathcal{V}$.

Furthermore, conisder the decomposition of a Hilbert space by tensor
products. Assume, for instance, two (isomorphic) Hilbert spaces $\mathcal{A}$
and $\mathcal{B}$ where $\mathcal{H}=\mathcal{A}\otimes\mathcal{B}$.
Then the full Hermitian Hamiltonian $\mathrm{\hat{H}}$ in $\mathcal{H}$
can be represented by

\begin{equation}
\mathrm{\hat{H}}=\mathrm{\hat{H}}_{\mathcal{A}}\otimes\hat{1}+\hat{1}\otimes\mathrm{\hat{H}}_{\mathcal{B}}+\Im_{\mathcal{AB}},\end{equation}
where $\Im_{\mathcal{AB}}$ denotes an arbitrary interaction. It is
straightforward then by (2) and (7) to show that

\begin{equation}
i\partial_{t}\left(\psi_{\mathcal{A}}\otimes\psi_{\mathcal{B}}\right)=\mathrm{\hat{H}}_{\mathcal{A}}\psi_{\mathcal{A}}\otimes\psi_{\mathcal{B}}+\psi_{\mathcal{A}}\otimes\mathrm{\hat{H}}_{\mathcal{B}}\psi_{\mathcal{B}}+\Im_{\mathcal{AB}}\left(\psi_{\mathcal{A}}\otimes\psi_{\mathcal{B}}\right).\end{equation}
Equation (8) does not itself represent a Schr\"odinger equation for
each partial Hamiltonian, unless one excludes the interaction $\Im_{\mathcal{AB}}$.
Thus, if a decomposition of the full Hermitian Hamiltonian is carried
out by tensor products such that the full Hamiltonian remains the
time generator so as to have that (2) fulfilled, then it follows that
(1) does not necessarily hold for every partial Hamiltonian, unless,
that is to say, one excludes interactions in a multi-particle system.

To conclude, the formalism of multi-time equations in quantum mechanics
{[}11{]} overlooks the class of physical systems for which Schr\"odinger
picture does not exist. The integrability condition (3) does not necessarily
imply that for an aribitrary decomposition of the Hilbert space (e.g.,
by direct sums) there exists a Schr\"odinger equation for every partial
Hamiltonian. Even if one considers particular tensor product decomposition
of a full Hermitian Hamiltonian, such that all partial Hamiltonians
remain self-adjoint, one has to conclude that the formalism excludes
and cannot cover interactions even in a non-relativistic multi-particle
quantum system. Ultimately it is notable that, due to the fact that
relative coordinates are not necessarily four-vectors, it is not possible
to simply include interactions in relativistc quantum mechanics (as
one does in non-relativisitc case) and get a manifestly Lorentz-invariant
theory. This subtle point was already elucidated by Eddington and
Dirac by their exchange notes through the Society {[}3,5,6{]}.

\subsection*{Acknowledgements}

I thank Nils Schopohl and Michael Benner for discussions.

\end{document}